# A strategy to tailor the mechanical and degradation properties of PCL-PEG-PCL based copolymers for biomedical application

Yu-Yao Liu [a, b], Juan Pedro Fernandez Blazquez [a], Guang-Zhong Yin [a, c], De-Yi Wang [a], Javier Llorca [a, b] * and Monica Echeverry-Rendón [a] *

[a] IMDEA Materials Institute, C/Eric Kandel 2, 28906, Getafe, Madrid, Spain

[b] Department of Materials Science, Polytechnic University of Madrid/Universidad Politécnica de Madrid, 28040 Madrid, Spain

[c] Universidad Francisco de Vitoria, Ctra. Pozuelo-Majadahonda Km 1,800, 28223, Pozuelo de Alarcón, Madrid, Spain

## Abstract

Biodegradable and biocompatible 3D printable biomaterials with tunable mechanical properties and degradation rate adapted to target tissues were urgently required to manufacture scaffolds for tissue regeneration. Herein, a strategy based on a series of copolymers are proposed where the mechanical and degradation properties can be optimized regarding the specific biological application. With this purpose, poly(ε-caprolactone)-poly(ethylene glycol)-poly(ε-caprolactone) (PCL-PEG-PCL, PCEC) triblock co-polymers with high molecular weight were synthesized by using PEG with a wide range of molecular weight (from 0.6 kg/mol to 35 kg/mol) as macroinitiators. PCEC copolymers exhibited tunable mechanical properties with an elastic modulus in the range 338-705 MPa and a degradation rate from 60% mass loss after 8 h to 70% mass loss after 23 days in accelerated tests, as well as excellent cytocompatibility and cell attachment after culture with mouse fibroblast L929 cells. The mechanisms responsible for these properties were ascertained by means of different techniques to ascertain the structure-property relationship in PCEC copolymers. Furthermore, it was shown that it is possible to manufacture PCEC scaffolds by 3D printing with excellent dimensional accuracy and controlled microporosity. This study provides a promising strategy to design, select, and fabricate copolymers with tunable mechanical properties and degradation rate for tissue engineering applications.

# 1. Introduction

Tissue engineering is currently considered an excellent approach to replace damaged tissue, and synthetic polymer scaffolds play a very important role in this task due to their similar structure and function to extracellular matrix (ECM)[1–3]. An ideal scaffold should physically and biologically mimic the structure and function of the target tissue. Specifically, the scaffold should provide adequate mechanical support for the damaged tissue and gradually degrade as new tissue grows. Besides, favorable biological properties are also required for tissue recovery [4–6]. However, the mismatch of mechanical properties and degradation rate [7,8], the presence of inflammatory degradation products [9,10], and difficulties to manufacture personalized scaffolds with precise pore interconnections [11,12] make it difficult to meet clinical use needs. Therefore, the development of materials with tailored mechanical properties and degradation rates that are suitable for different tissues and that can be used to fabricate customized and personalized scaffolds remains a significant challenge.

The emergence of 3D printing technologies provides broad prospects for the manufacture of personalized scaffolds for tissue engineering, since it can accurately and quickly fabricate scaffolds with precise geometry and high pore connectivity, similar to native tissues [13–17]. Among them, extrusion-based 3D printing is a common strategy because of low cost, simple and eco-friendly fabrication, high availability of materials (metals, ceramics, polymers, hydrogels and their composites) and diverse raw material to process (wires, pellets, and powders) [18–20]. In this context, some synthetic biodegradable polymers have attracted considerable attention over the past decades since the present high biocompatibility, ease of processing, as well as tailored mechanical properties and degradation rates [21–26]. For instance, poly($\varepsilon$-caprolactone) (PCL), approved by the Food and Drug Administration (FDA), is widely used in bone regeneration because of its mechanical properties, good biocompatibility, and easy processing due to the low melting temperature of ~ 60 °C[27–31]. However, PCL has a hydrophobic surface, which affects the degradation process that is reported to be between 2-3 years *in vivo*. The poor wettability also has a direct influence on the cell attachment and proliferation resulting in certain limitations in healthcare and medical applications [32–34].

As a nontoxic and water-soluble polymer, poly (ethylene glycol) (PEG) is often used for copolymerization or blending with PCL matrix to improve the hydrophilicity of PCL-based 3D printed scaffolds, and some interesting results have been achieved [35–37]. Among them, PCL-PEG-PCL (PCEC) triblock copolymers synthesized from ring-opening polymerization (ROP) are attractive in tissue engineering due to the controlled synthesis and the flexibility of compositions, which allows to tailor the mechanical properties and degradation rate of PCL-based scaffolds [38–41]. However, most studies have focused on the regulation of PCEC copolymer composition in a narrow range and low molecular weights, because high molecular weights will bring high viscosities which are not conducive to 3D printing, as well as low degradation rates which are not compatible to tissue regeneration, thus limited the development of researches. In addition, to our best knowledge, the structure-property relationships of PCEC copolymers from the viewpoint of degradation rate, biological performance, and fabrication for tissue engineering applications have not been systematically explored, especially for macroinitiators of PEG with high molecular weight of 20k g/mol and 35k g/mol. Therefore, systematic studies are still necessary to design and select PCEC copolymers and scaffolds that can be used for tissue engineering in different clinical applications.

Herein, this study aims to produce and characterize biocompatible and biodegradable PCEC copolymers with high molecular weight using macroinitiators from low molecular weight PEG (0.6k g/mol) to high molecular weight PEG (35k g/mol). They are expected to cover a wide range of mechanical properties, degradation rates and biological performance, so they can be used in different applications. The composition, thermal properties, crystallization behavior, mechanical properties, degradation rate, and biological properties are studied systematically to define the structure-property relationships of PCEC copolymers. Finally, they were used to manufacture scaffolds via a pellet-based screw-driven 3D printer, so the potential for personalized medicine is demonstrated.

# 2. Materials and Methods

## 2.1 Synthesis of PCL and PCEC copolymers

As shown in **Figure 1a**, PCEC copolymers were synthesized by ring-opening polymerization (ROP) of Ɛ-CL and PEG as previously described[42]. Typically, 0.2 mmol polyethylene glycol (PEG, Sigma-Aldrich, Mn= 0.6k, 2k, 6k, 20k and 35k g/mol, dried before use) and 0.175 mol Ɛ-caprolactone (Ɛ-CL, Sigma-Aldrich, 97%, distilled) were weighed into a Schlenk flask, and the catalyst of tin (II) 2-ethylhexanoate ($Sn(Oct)_2$, Sigma-Aldrich, 92.5-100.0%) with a 0.5 wt% ratio was dropped into the mixture. Then, the branch pipe on the flask was connected to a vacuum pump for 10 min to remove the residual moisture and oxygen in the flask. After that, the flask was immersed in an oil bath at 110 °C under stirring for 24 h. Then, the product was dissolved in dichloromethane ($CH_2Cl_2$, TCI EUROPA N.V. Belgium, 99.0%, anhydrous) and precipitated into petroleum ether (Sigma-Aldrich, boiling point ≥ 90% 40-60 °C) three times to purify the product. Because the catalyst $Sn(Oct)_2$ can also be dissolved in petroleum ether, it was removed during the dissolution and precipitation. After filtration and drying in a vacuum oven at 40 °C for 24 h, white block products were obtained (yield: > 95%). According to the molecular weight of initiators, the obtained samples were named by $PCE_{0.6k}C$, $PCE_{2k}C$, $PCE_{6k}C$, $PCE_{20k}C$ and $PCE_{35k}C$, respectively. PCL was synthesized as above from ethylene glycol (EG, Sigma-Aldrich, 99.8%, anhydrous) and Ɛ-CL at the same ratio.

## 2.2 Preparation of polymer film

PCL and PCEC copolymer samples (~ 7 g) were transferred to a mold (100 × 100 × 0.5 mm$^3$) and processed by hot-pressing at 120 °C for 10 min under a pressure of 20 MPa to obtain regular polymer films for characterizations. Surface morphology of the polymer films were assessed by using atomic force microscopy (AFM) (XE-150, Park Systems). Areas of 40×40 μm were scanned in contact mode at a rate of 0.3 Hz. Average roughness (Ra) were measured and recorded from the surface.

## 2.3 Characterization

### 2.3.1 Composition and structure

The composition and structure of PCL and PCEC copolymers was analyzed by $^1$H-NMR, $^{13}$C-NMR (AVANCE III 400, Bruker, USA) and Fourier transform infrared spectroscopy (FT-IR, Thermo Fisher Nicolet 5700 spectrometer). $^1$H NMR and $^{13}$C-NMR analysis spectrum were measured at 400 MHz using deuterated chloroform as the solvent, while FT-IR was performed under an ATR mode.

Then, the molecular weight and the poly-dispersity index of PCL and PCEC copolymers were obtained from a gel permeation chromatographer (GPC, Waters1525, Spain) that equipped with a Waters 2424 refractive index detector and a Waters 2489 UV detector, this process was carried out at 25 °C. Monodisperse polystyrene (Sigma-Aldrich, Germany) was used as the standard sample, and tetrahydrofuran (THF, Sigma-Aldrich, Germany) was used as the solvent.

Additionally, dynamic mechanical analysis (DMA), differential scanning calorimetry (DSC) and X-ray diffraction analysis (XRD) were performed to investigate the thermal properties and crystallization behavior of PCL and PCEC copolymers. DMA was studied by using a TA- Q800 machine, and the samples were test from -110 °C to 40 °C under with a heating rate of 3 °C/min and a frequency of 1 Hz. DSC was performed for each film (5-10 mg) using a TA-Q200 machine in an $N_2$ atmosphere (50 mL min$^{-1}$). Samples were tested from -40 °C to 100 °C and kept for 1 min to remove the thermal history, then quenched to -40 °C at a cooling rate of 10 °C/min, and after isothermal equilibrium for 1 min, heated to 100 °C at a rate of 10 °C/min. XRD was conducted on an X-ray diffractometer (PANalytical B.V., Netherlands) with Ni-filtered Cu Kα radiation from 10° to 40°.

Rheology properties were explored on a rheometer (Anton Paar MCR 702e, Austria) by using disk samples with a diameter of 25 mm and a thickness of 1 mm at 110 °C under a shear stain of 3% in a frequency range of 0.1-50 rad/s. The storage modulus (G'), loss modulus (G'') and complex viscosity (η*) were recorded during the experiment.

### 2.3.2 Mechanical characterization

PCL and PCEC copolymer films were cut into dog-bone shape with dimensions of 75 mm × 4 mm × 0.5 mm, following ISO 527/2 5A (1996) standard. A universal tensile testing machine (Instron 5966, USA) was utilized to carry out the tensile tests with a 500 N load cell and an initial length of 40 mm at a speed of 10 mm/min. A minimum of 6 specimens were tested for each sample.

### 2.3.3 Hydrophilicity test

The water contact angle of PCL and PCEC copolymers was measured by a Drop Shape Analyzer (DSA25S, KRÜSS). A drop of water (3 µL) was served on the surface of samples and the image was immediately recorded. The behavior of the drop and the material was recorded for 5 seconds, and the average and the standard deviation were obtained. A minimum of 3 specimens were tested for each sample.

### 2.4 Degradation test

Accelerated degradation tests were carried out in 5 M sodium hydroxide (NaOH, Sigma-Aldrich, Germany) solution at 37 °C to explore the degradation behavior. Briefly, PCL and PCEC copolymer films with a diameter of 15 mm × 5 mm × 0.5 mm were placed into glass vials, respectively. Then 2 mL of 5 M NaOH solution was added to each vial to immerse the whole film. After tightly sealed, the vials were placed into an oven with a temperature of 37 °C. Samples were checked every 2 or 3 days to monitor the change in mass. To this end, samples were removed, washed with deionized water, and dried in a fume hood for 24 h. Then samples were weighed every time and put back to the glass vials. The process was repeated until the samples completely lost their structural integrity. A least 3 pieces of each sample were performed at the same time to obtain the average and standard deviation.

### 2.5 Cytocompatibility test

2.5.1 Cell culture

A mouse fibroblast L929 cell line (ATCC-CRL-2593) was used for the cytocompatibility tests. Cells were cultured in a complete medium consisted of high-glucose Dulbecco's modified eagle medium (DMEM, Gibco, USA), fetal bovine serum (FBS, Corning, USA), penicillin (100 units/mL, Invitrogen, USA) and streptomycin (100 µg/mL, Invitrogen, USA). Cells were used at 80 % of confluence and incubated at 37 °C with an atmosphere of 95 % air and 5 % $CO_2$.

2.5.2 Indirect tests

Extracts of the PCL and PCEC copolymers were obtained by immersion of polymer films in complete culture medium with a ratio of 3 $cm^2$/mL and incubated at 37 °C for 72 h. Then, the samples were removed and extracts were used for the cytocompatibility assay. To this end, L929 cells were seeded with a cell density of $1 \times 10^4$ cells/cm$^2$ and incubated at 37 °C for 24 h. Then, the culture medium was removed and replaced with diluted extracts (100%, 50%, 25%, 12.5%, 6.25% and 3.125%) and incubated for another 24 h. The mitochondrial activity was measured by adding thiazolyl blue tetrazolium bromide (MTT, Sigma-Aldrich, Germany) at a concentration of 5 mg/ml with a ratio of 10 µL by each 100 µL of complete culture medium. Afterwards, the cells were incubated under the same conditions for another 3 h and 100 µL dimethyl sulfoxide (DMSO, Sigma-Aldrich, Germany) was added in each well to dissolve the formazan crystals produced. Finally, the absorbance was measured in a microplate reader (Tecan Infinite M Plex, Switzerland) at a 570 nm wavelength. Cells without extract but with complete culture medium were used as the control. The mitochondrial activity was reported after normalize the optical density (OD) of the cells exposed to the extracts with respect the control (mitochondria activity (%) = (OD of sample / OD of positive control) × 100 %).

2.5.3 Direct tests

Mouse fibroblast L929 cells were seeded at a cell density of $1 \times 10^4$ cells/cm$^2$ directly on the surface of the samples and incubated at 37 °C for 24 h. Afterwards, cells were fixed with paraformaldehyde (PFA 4%, Thermo Fisher Scientific, USA) at 37 °C for 25 min. Cell morphology was studied for immunofluorescent staining. After fixation, cells were treated with 0.1% of triton® X-100 (Thermo Scientific, USA) for 15 min and washed several times with Dulbecco's phosphate buffered saline (D-PBS, Corning, USA). Afterwards, cells were stained with Alexa Fluor 488 phalloidin (Invitrogen, USA) at a concentration of 1:300 to observe the cytoskeleton, and with DAPI (Invitrogen, USA) at a concentration of 1:1000 for the identification of the nucleus. After incubation under dark at room temperature for 1 h, the morphology of cells was observed under confocal microscopy (Olympus FV3000, Japan).

The cell-material interaction was studied on cells seeded directly in the sample in a scanning electron microscope (SEM, FEI Helios NanoLab 600i, USA). After fixation with PFA 4%, samples were dehydrated by immersion in increasing concentrations of ethanol in distilled water (30%, 50%, 70%, 90% and 100%) for 10 min each. Then, the process was repeated by increasing 1,1,1,3,3,3-hexamethyldisilazane (HMDS, TCI EUROPE N.V., Belgium) in ethanol at concentration of 33%, 50%, 66% and 100% for 10 min each. Samples were dried overnight and then observed in the SEM at magnifications of 500 x and 2500 x.

**2.6 Fabrication of 3D-printed scaffolds for biomedical applications**

A pellet-based 3D printer with a screw-assistant system (Direct 3D, Italy) was used to manufacture scaffolds. Firstly, the PCL and PCEC copolymers were dissolved in $CH_2Cl_2$ solution and cast into flat containers to dry into thin films in the vacuum oven at 40 °C overnight, respectively. Then, films were cut into small pieces for the 3D printing. The printing task was conducted in the following parameters; the nozzle of the printer has a diameter of 500 μm, the extrusion temperature was 110 °C, the experimental temperature and printing bed temperature was 25 °C, the layer thickness 500 μm, the printing speed was 5-10 mm/s with screwing flow percentages of 2000-8000%, in which the difference depended on the rheological properties of each sample).

Scaffolds with different features were designed. They include 2-layer sheets (strut thickness of 500 μm, pore size of 1000 μm, and 0°/90° lay-down pattern) of 30 mm × 30 mm × 1 mm, 10-layer scaffolds (strut thickness of 500 μm, pore size of 1000 μm, and 0°/90° lay-down pattern) of 15 mm × 15 mm × 5 mm, and 10-layer scaffolds (strut thickness of 500 μm, pore size of 500 μm, and 0°/90° lay-down pattern) of 15 mm × 15 mm × 5 mm. The microstructure of 3D-printed scaffolds was observed in an optical microscope (Olympus SC50, Japan) to assess the accuracy of the 3D printing process.

# 3. Results and discussion

## 3.1 Synthesis and structure of PCL and PCEC copolymers

PCL and a series of PCEC block copolymers with different PEG contents were successfully synthesized by the ring-opening reaction of ε-CL as the monomer, and EG or PEG as initiators using $Sn(Oct)_2$ as the catalyst (**Figure 1a**). PEG0.6k, PEG2k, PEG6k, PEG20k, and PEG35k were used as macroinitiators to assess the structure-property relationships as a function of PEG content on PCEC copolymers. $^1$H-NMR, $^{13}$C-NMR, FT-IR, and GPC analysis were conducted to confirm the structure, composition and molecular weight of the synthesized block copolymers. The assumed schematic diagram of blocks in PCEC copolymers is shown in **Figure 1b**.

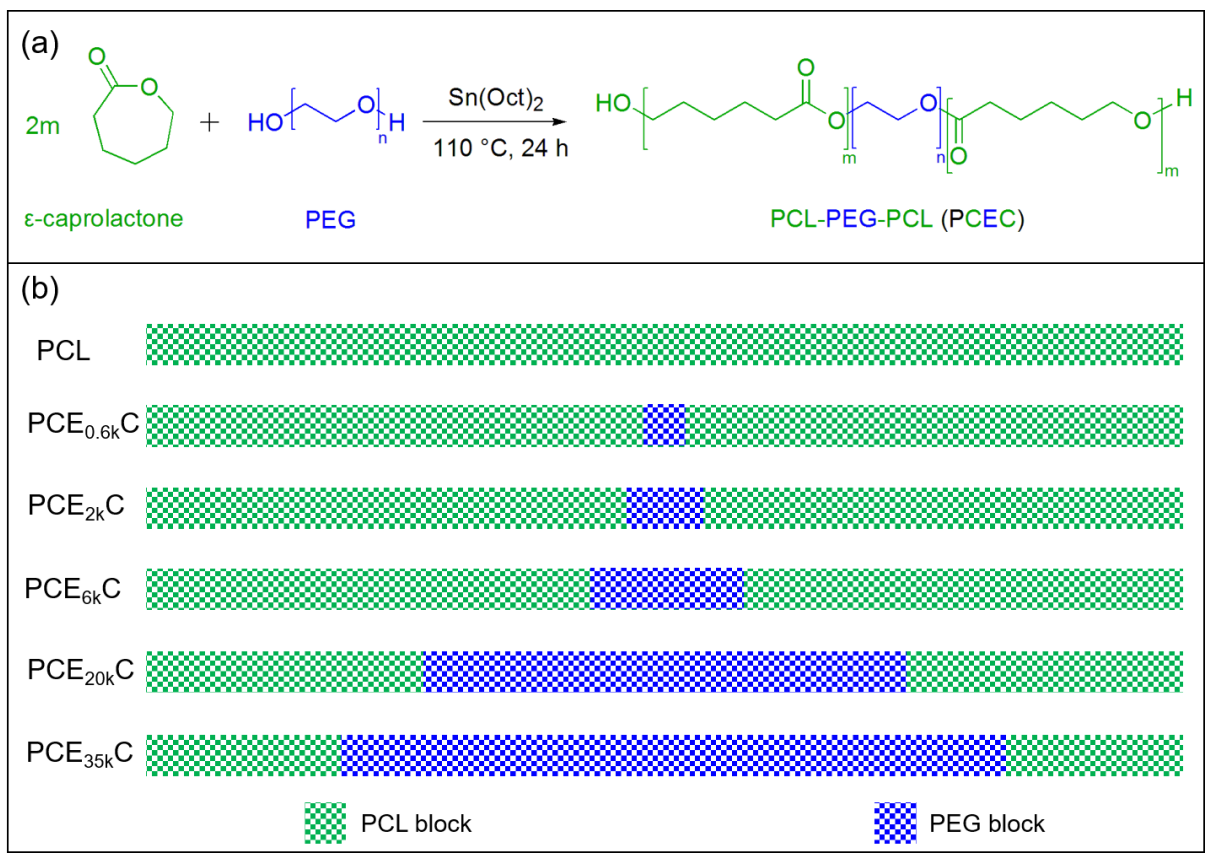

**Figure 1.** (a) Synthesis route and (b) schematic diagram of the composition of PCL and PCEC copolymers.

The structures and characteristic $^1$H-NMR spectra of PCL and PCEC copolymers after purification are shown in **Figures 2a, 2b, and 2c**. The peaks at 4.05 ppm (peak e), 2.29 ppm (peak a), 1.59 to 1.68 ppm (peak b and d), and 1.33 to 1.41 ppm (peak c) in **Figure 2a** correspond to the methylene protons of the PCL blocks. In addition to the appearance of PCL block peaks, a typical peak appears in PCEC copolymers curves at 3.65 ppm (peak h), corresponding to the methylene protons of the PEG block [43] (**Figures 2b and 2c**). Furthermore, the appearance of characteristic peaks of the PCL block and PEG block in the $^{13}$C-NMR and FT-IR spectrum also confirms the composition of the PEG block and PCL block (**Figure S1 and Figure S2**). All this evidence indicates that PCEC copolymers made up of PEG and PCL blocks were successfully synthesized. In addition, the proportions of PEG block and PCL blocks in the synthesized PCEC copolymers were also different due to the different molecular weights of the macroinitiator PEG. It can be seen in **Figure 2c** that the peaks of the PEG block became stronger with the increase in the molecular weight of PEG while the peaks of the PCL block became weaker, which is because the PEG content in the PCEC copolymers increased gradually.

**Table 1.** Composition and molecular weight of PCL and PCEC copolymers

| Sample | PCL/PEG block ratio [a] | $M_n$ [b] (g/mol) | $M_w$ [c] (g/mol) | PDI [d] |
|---|---|---|---|---|
| PCL | 121.9 | 65 196 | 94 194 | 1.44 |
| PCE$_{0.6k}$C | 30.3 | 73 275 | 105 412 | 1.43 |
| PCE$_{2k}$C | 9.8 | 89 457 | 121 256 | 1.39 |
| PCE$_{6k}$C | 3.2 | 79 293 | 111 810 | 1.41 |
| PCE$_{20k}$C | 0.96 | 61 879 | 79 667 | 1.28 |
| PCE$_{35k}$C | 0.55 | 72 047 | 89 554 | 1.24 |

[a] PCL/PEG block ratio calculated from the peak area ratio of peak e and h in $^1$H-NMR
[b] Number-average molecular weight ($M_n$) obtained from GPC
[c] Weight-average molecular weight ($M_w$) obtained from GPC
[d] Polydispersity index (PDI, $M_w/M_n$) obtained from GPC

In order to figure out the specific proportions, PCL / PEG block ratios in PCEC copolymers were obtained from the calculation of the peak area ratio of **e** and **h** in $^1$H-NMR (2m/n, where n refers to the polymerization degree (DP) of PEG block and 2m refers to the total DP of PCL blocks) (**Figure 1a**), and the results are depicted in **Table 1**. As the molecular weight of PEG increases, the ratio gradually decreases, corresponding to the increase of the PEG block and the decrease of the PCL block. It is worth noticing that the ratio was close to 1 in PCE$_{20k}$C, indicating that the polymerization degree (DP) of these two blocks was nearly equal in the copolymer. Besides, the DP of PCL block was smaller than that of PEG block (2m/n ratio <1) in PCE$_{35k}$C, demonstrating that the PEG block content occupies a significant part in copolymer. Therefore, these copolymers can be classified into three composition types according to the block content: (1) PEG < PCL, (2)

PEG ≈ PCL, and (3) PEG >PCL, as shown in **Figure 1b**. These different compositions will significantly affect the properties of the copolymers, as it will be discussed in detail below.

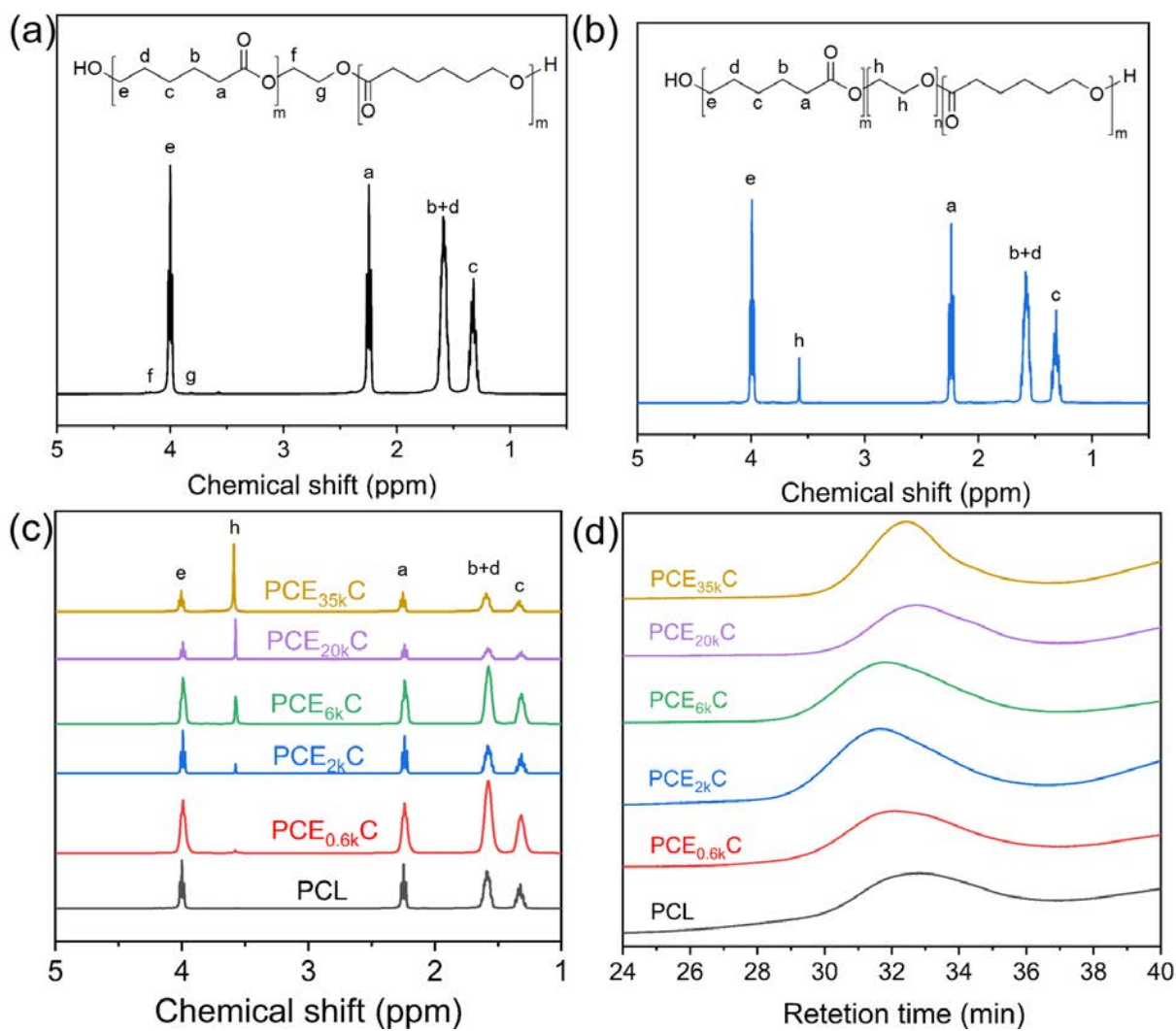

**Figure 2.** The $^1$H-NMR spectra of (a) PCL, (b) PCE$_{2k}$C, (c) PCL and PCEC copolymers, and (d) GPC curves of PCL and PCEC copolymers

GPC measurements were performed to determine the molecular weight of PCL and PCEC copolymers and the results are presented in **Table 1**. As shown in **Figure 2d**, all the relative curves of PCL and PCEC copolymers show a single peak of relative molecular weight. The PCL and PCEC copolymers exhibit high molecular weights within the range of 60k g/mol to 90k g/mol. These results prove that the synthesis is controlled and these PCL and PCEC copolymers can be used for comparison.

## 3.2 Thermal properties and crystalline behavior

Thermal properties and crystalline behavior are critical in tissue engineering because they can significantly affect the mechanical properties and degradation rate. The thermal properties, crystalline behavior and rheological properties of PCL and PCEC copolymers were studied by DSC, DMA, XRD and rheology (**Figure 3 and Figure S4**), and the related data are listed in **Table 2**.

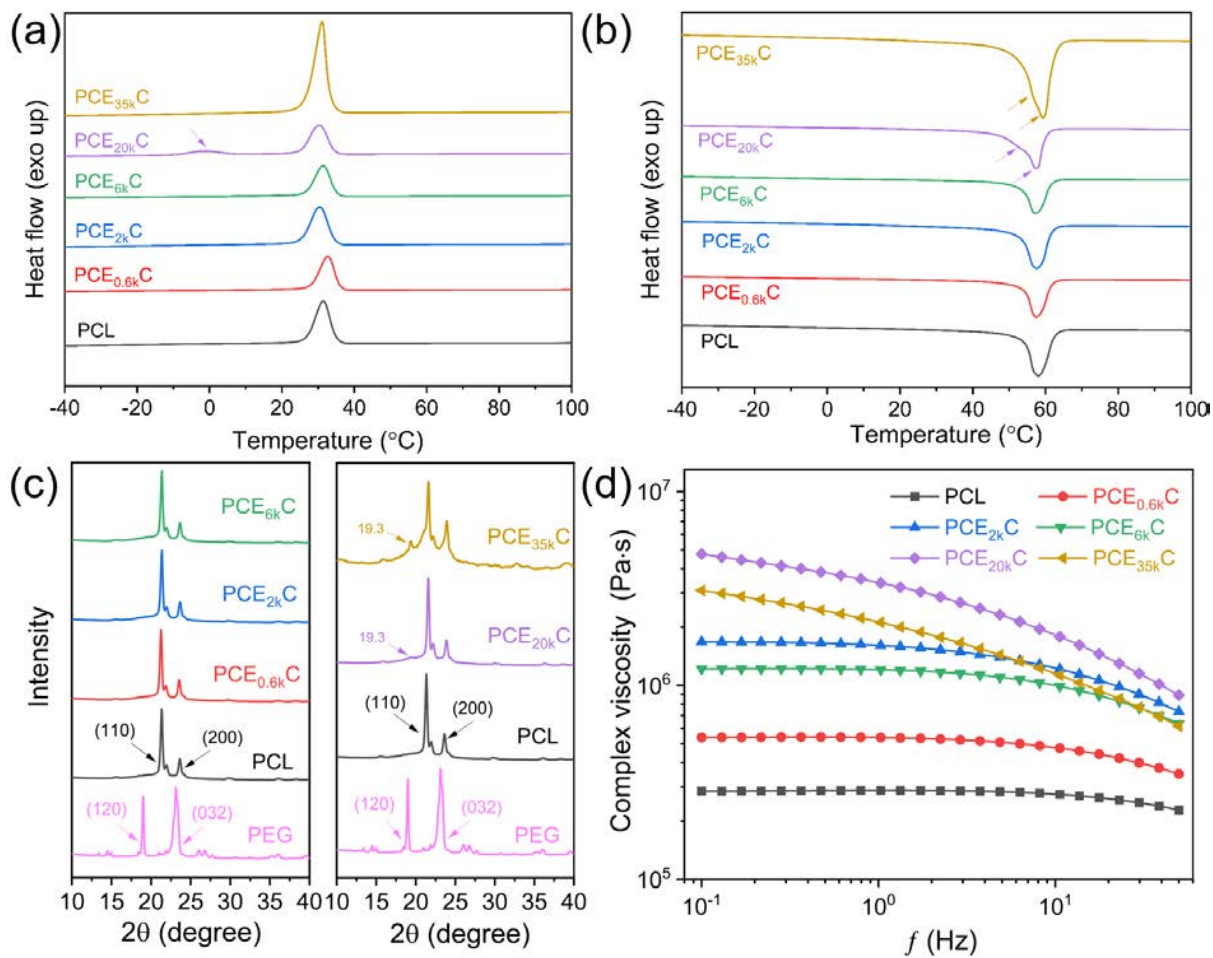

**Figure 3.** DSC curves of PCL and PCEC copolymers after removing the thermal history: (a) crystallization process and (b) melting process, (c) XRD spectra of PCL and PCEC copolymers and (d) complex viscosity of PCL and PCEC copolymers in rheology

As shown in the DSC curves, $PCE_{0.6k}C$, $PCE_{2k}C$, and $PCE_{6k}C$ display a single crystallization temperature ($T_c$) at around 30 °C corresponding to the PCL block with a $T_c$ of 31.4 °C (**Figure 3a**). Due to the small DP of the PEG blocks in these copolymers, the mobility of PEG blocks is affected by the crystallization of PCL blocks, thus leading to a single peak related to the PCL block. Besides, the PCL block content in the copolymer decreases with the increase of the PEG block content, leading to a reduction in the crystallization enthalpy ($\Delta H_c$) of PCL from 64.0 to 57.8 J/g. The same tendency can also be observed in the melting process, as shown in **Figure 3b and Table 2**. However, when the PEG block content increases (PEG ≈ PCL), the crystallization of PEG begins to appear, resulting in two crystallization peaks in PCE$_{20k}$C, where the one at 30.2 °C corresponds to the PCL block and the other at -2.1°C stands for the crystallization of the PEG block. This result can also be confirmed from the overlapping melting peaks and a higher $\Delta H_m$ (melting enthalpy, 75.0 J/g). However, the actual $T_c$ of PEG20k was around 45 °C according to DSC (**Figure S3**). This reduced temperature comes from the crystallization of the PCL block at high temperature, which affects the mobility of the PEG block, leading to the difficulty in PEG crystallization. Conversely, when PEG > PCL, PEG block crystallization dominates and appears as a sharp peak similar to the crystallization peak of the PEG block (**Figure S3**) in PCE$_{35k}$C, but with a higher $\Delta H_c$ of 79.7 J/g due to the joint contribution of PEG and PCL blocks. The overlap also appears in the melting process of PCE$_{35k}$C, which exhibits a higher $\Delta H_m$ = 89.0 J/g. Besides, it also can be confirmed by the change of glass transition temperature ($T_g$) from DMA (**Figure S4** and **Table 2**). When PEG < PCL, the crystallization enthalpy of the copolymers decreased with the increase of PEG block content, resulting in the decrease in $T_g$. While when the crystallization of PEG begins to appear, the $T_g$ of the copolymer increases. It is because the $T_g$ of the copolymer decreases as the crystallinity of the copolymer decreases, which is consistent to the results of DSC.

Table 2. Thermal properties of PCL and PCEC copolymers

| Sample | $T_g$(°C) | $T_c$ (°C) | $T_m$ (°C) | $\Delta H_c$ (J/g) | $\Delta H_m$ (J/g) |
|---|---|---|---|---|---|
| PCL | -38.3 | 31.4 | 58.1 | 64.0 | 70.5 |
| PCE$_{0.6k}$C | -39.2 | 32.7 | 57.4 | 62.6 | 67.6 |
| PCE$_{2k}$C | -43.6 | 30.5 | 57.5 | 59.6 | 66.2 |
| PCE$_{6k}$C | -45.8 | 31.3 | 57.3 | 57.8 | 62.1 |
| PCE$_{20k}$C | -39.4 | -2.1, 30.2 | 53.3, 57.3 | 10.3, 53.3 | 75.0 |
| PCE$_{35k}$C | -41.4 | 31.0 | 59.4 | 79.7 | 89.0 |

To gain insight into the crystallization behavior of these block copolymers, X-ray diffraction (XRD) was performed. As shown in **Figure 3c**, the peaks at 21.3° and 23.6° corresponded to (110) and (200) crystal planes of PCL block, while the peak at 19.0° and 23.1° related to (120) and (032) crystal planes of PEG block [44]. Only peaks corresponding to the PCL block appear when PEG < PCL, indicating the crystallization of PEG block with lower DP is restricted by the PCL block crystallization. A small peak corresponding to the PEG block crystallization appears at 19.3° when PEG ≈ PCL, because of the increased DP of PEG block in $PCE_{20k}C$. More importantly, both peaks corresponding to the PEG block and the PCL block appear clearly in $PCE_{35k}C$ when PEG > PCL, providing a strong evidence of the joint contribution of PEG and PCL blocks and the overlap phenomenon observed in DSC.

For extrusion-based 3D printing, the rheological behavior of the material is a key factor as it greatly affects printability. In particular, the shear thinning behavior is beneficial to 3D printing, because the low viscosity under high-frequency shear is conducive to the extrusion of the material which prevents the nozzle clogging; while the high viscosity under low-frequency shear is beneficial to maintain the shape after deposition and preventing microstructure collapse. The rheology of PCL and PCEC copolymers was explored and the results are plotted in **Figure 3d**. PCL shows a typical Newtonian behavior that was frequency-independent in the low frequency region. It was interesting to see that the PCEC copolymers show a PCL Newtonian behavior when PEG < PCL, while they exhibit an obvious shear-thinning phenomenon when PEG ≥ PCL, which is the key factor for extrusion-based 3D printing. This special performance comes from the increase of the flexible PEG blocks, which improve the flexibility of the copolymer chain, leading to decreased viscosities at high shear frequency. Moreover, higher moduli G' of $PCE_{20k}C$ and $PCE_{35k}C$ appear at low shear frequency, making them more suitable to obtain a stable microstructure after deposition (**Figure S5**). These results demonstrate that the rheological properties of PCEC copolymers can be optimized for the 3D printing by carefully controlling the composition. Furthermore, it also provides an effective method to improve the rheological properties of PCL-based polymers without reducing the molecular weight or blending with other materials.

### 3.3 Mechanical properties

Tensile tests were carried out to ascertain the mechanical properties of PCL and PCEC copolymers. The stress-strain curves and tensile properties are plotted in **Figure 4** and the corresponding values of the elastic modulus, strength and elongation at fracture are given in **Table S1**. All copolymers showed an initial elastic region, followed by plastic deformation after the yield point, leading to a large elongation to fracture (**Figure 4a**). The elastic modulus of PCEC copolymers decreases while the elongation at fracture increases with the increase of PEG block content if PEG ≤ PCL (**Figure 4b**). This trend behavior comes about because PCL block dominates in the crystallization of theses copolymers due to the lower DP of PEG block, the reduced PCL block content decreases the crystallinity, resulting in lower elastic modulus. However, the crystallization of the PEG block dominates when PEG > PCL, and the combination of the PEG and PCL blocks leads to a higher crystallinity, resulting in higher elastic modulus. Although the mechanical properties are usually related to the molecular weight of polymers, it can be seen from **Table 1** that the molecular weights of PCL and PCEC are similar, which indicates that the mechanical properties of PCEC copolymers are mainly influenced by different crystallization behaviors. It is interesting to notice that the elastic modulus of PCEC copolymers can be controlled in a wide range from 338 MPa to 705 MPa by adjusting the molecular weight of macroinitiator PEG, without reducing the molecular weight and elongation at fracture. Thus, these PCEC copolymers have a large potential to manufacture scaffolds for cancellous bone (~ 400 MPa), tendon (~ 560 MPa) and collateral ligament (~ 366 MPa) in the tissue engineering, taking into account their large elongation at fracture.

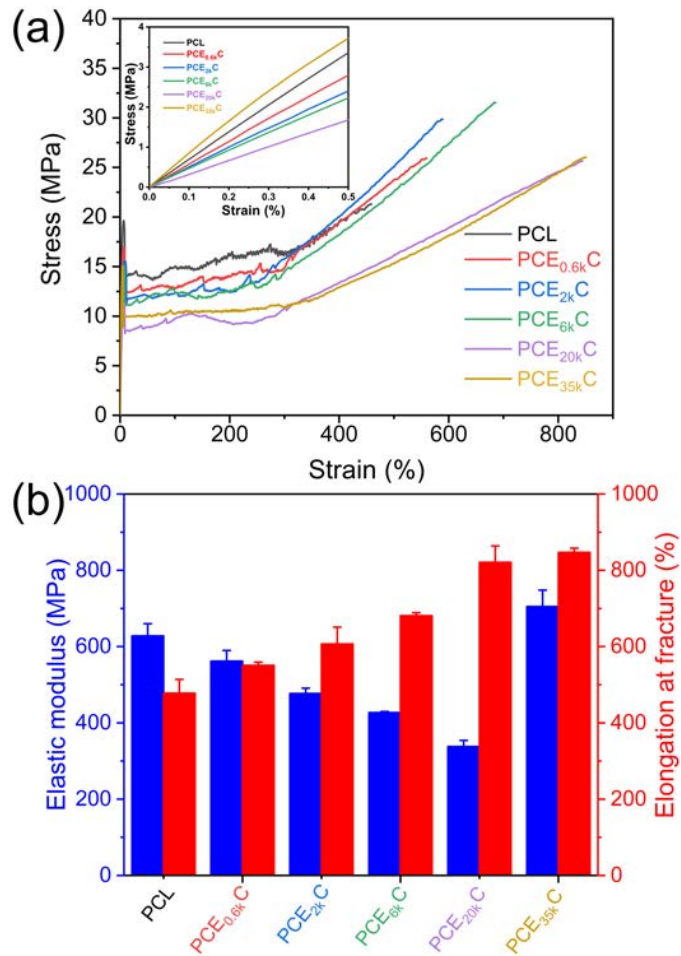

**Figure 4.** (a) Typical stress-strain curves and (b) tensile properties of PCL and PCEC copolymers

### 3.4 Wettability

Water contact angles were explored before degradation tests to determine the surface hydrophilicity of PCEC copolymers, as it directly affects the degradation rate and cell behavior. PCL, PCE$_{0.6k}$C and PCE$_{2k}$C showed similar contact angles of 84.2°, 87.3° and 87.1°, respectively, representing a hydrophilic behavior (contact angle lower than 90°) (**Figure 5a**). The increase of the PEG block content in this range was insufficient to increase the hydrophilicity significantly. The contact angle begins to drop from PCE$_{6k}$C (73.4°), and when PEG ≥ PCL, PCE$_{20k}$C and PCE$_{35k}$C show excellent hydrophilicity, with contact angles of 45.2° and 65.4°, respectively. This is due to the increase of the presence of the highly hydrophilic PEG blocks. Notably, PCE$_{35k}$C with higher PEG content shows a contact angle higher than that of PCE$_{20k}$C due to the joint contribution of crystallization discussed before, which makes it more difficult for water to diffuse on the surface. These results indicate that the hydrophilicity can be changed in a wide range by adjusting the PEG block content, providing a useful strategy to tailor the degradation rate and cell behavior.

### 3.5 *In vitro* accelerated degradation properties

The degradation rate is a vital factor for biomedical applications as it will directly affect the *in vivo* performance of the scaffolds. The degradation mechanism of PCL under physiological conditions (water or PBS solution) is bulk erosion. The sample first experiences a reduction in molecular weight, which later leads to the loss of mass. However, the degradation rate is not constant in bulk erosion, because of the complexity of degradation process [45]. However, the degradation mechanism of PCL in the alkaline environment with high pH is surface erosion. Under these conditions, the mass loss appears in the initial stage and the degradation rate are linear [46,47]. This constant degradation rate facilitates the comparison of the degradation among different materials. Therefore, accelerated degradation tests were performed by immersing PCEC copolymer films in 5M NaOH solution (pH = 13) at 37 ºC in this study. The degradation was monitored by measuring the mass residue over time until structural integrity was lost (**Figure 5b**). When PEG < PCL, PCL, PCE$_{0.6k}$C and PCE$_{2k}$C show similar degradation rates with ~70% mass loss retained structural integrity after 23 days, while PCE$_{6k}$C shows a faster degradation rate, with about 90% mass loss after 20 days and loss of structural integrity after 23 days. Surprisingly, when PEG ≥ PCL, PCE$_{20k}$C and PCE$_{35k}$C exhibit extremely fast degradation rates and they were fully degraded within 1 day, so that samples were checked every 2 h. PCE$_{20k}$C and PCE$_{35k}$C show ~60% mass loss after 8 h and lost their structural integrity after 10 h (**Figure 5c**). These phenomena were consistent to the results of hydrophilicity. During the degradation process, the hydrophilic PEG blocks were more likely to be released into the solution, leading to the formation of holes, which facilitate the ingress of water and accelerate the hydrolysis of the PCL block. However, the PEG blocks did not completely disappear due to surface erosion, but were gradually degraded as the thickness decreased. Therefore, it demonstrates that the apparent degradation rate of PCEC copolymers can be significantly affected by the PEG block content over a wide range, which broadens the design space of biomaterials with controllable degradation rates.

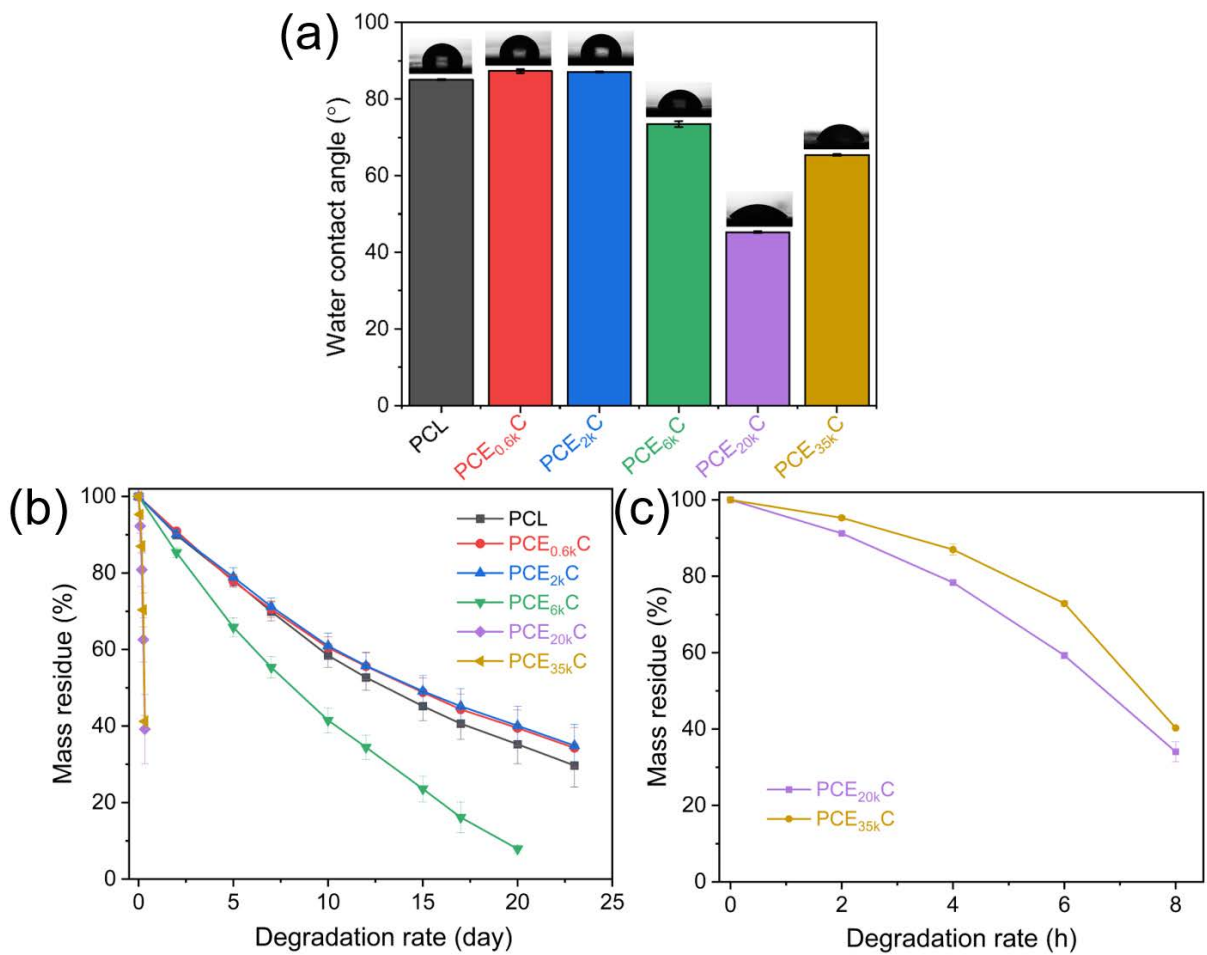

**Figure 5.** (a) Water contact angles of PCL and PCEC copolymers. (b) Mass residue of PCL and PCEC copolymers after immersion in NaOH solution for 23 days. (c) Mass residue of $PCE_{20k}C$ and $PCE_{35k}C$ after immersion in NaOH solution for 8 hours.

### 3.5 *In vitro* biological studies

Cytotoxicity of PCL and PCEC copolymers was tested by MTT assay. To this end, different diluted extracts (100%, 50%, 25%, 12.5%, 6.25%, and 3.125%) were used to incubate L929 cells at a concentration of $1 \times 10^4$ cells/cm$^2$ for 24 h. The mitochondrial activity of PCL and PCEC copolymers was normalized by the obtained value of the control and the relationship between the mitochondrial activity at function of the dilution is plotted in **Figure 6** for all copolymers. PCL and PCEC copolymers present a mitochondrial activity above 70%, which is the lowest limit to consider the material cytotoxic according to ISO10993-5. Thus, all copolymer present excellent biocompatibility, and there was no significant difference between them, which indicates that the composition of PCEC copolymer has no effect on the material cytotoxicity.

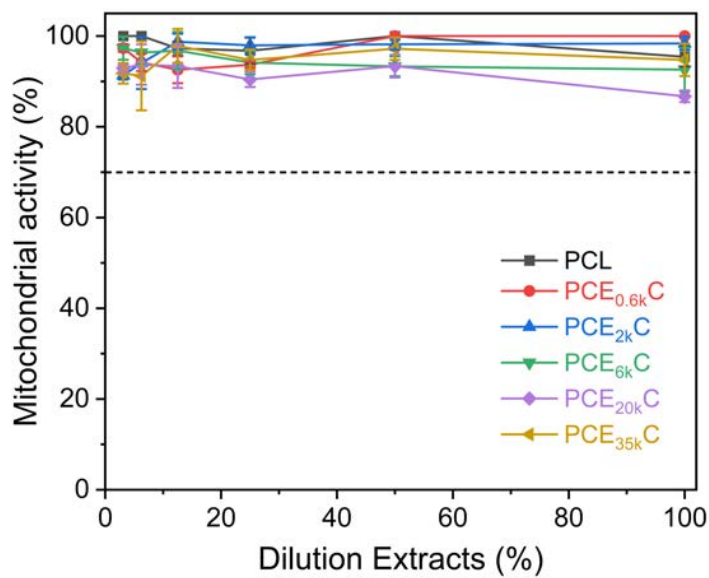

**Figure 6.** Mitochondrial activity of the L929 cells after 72 h of incubation in the extracts obtained from PCL and PCEC copolymers immersed in culture medium.

The homogeneity of the polymer films was observed by AFM. As shown in **Figure S6**, all samples present smooth surfaces, and the differences in surface morphology come from the roughness of the hot-pressed plate and possibility from the removal of material during the release of the hot-pressed films. However, the overall roughness is relatively homogeneous, and the average roughness (Ra) is between 0.35-1.76 μm, which does not affect the proliferation and growth of cells with size of 15-30 μm. The results of the direct interaction between PCL or PCEC copolymers and L929 cells after 24 h of incubation is depicted in **Figure 7**. Cells are viable on both PCL and PCEC copolymers and they display the typical morphology of L929 cells, indicating that these materials were biocompatible. Besides, it can be clearly seen from the SEM images that the cells on all samples exhibited extended cytoplasm and clear pseudopodia (**Figure S7**). Moreover, the quantity and quality of cell spreading seems to be improved with the increase in PEG content, and more cells with extended cytoplasmic projections appear, indicating enhanced cell adhesion, especially starting from PCE$_{6k}$C. This may be because the increase of PEG improves the hydrophilicity and surface energy, resulting in easier cell spreading and migrating on the material. These results confirm that the increased PEG content has an indirect effect on cell attachment and proliferation in PCEC copolymers.

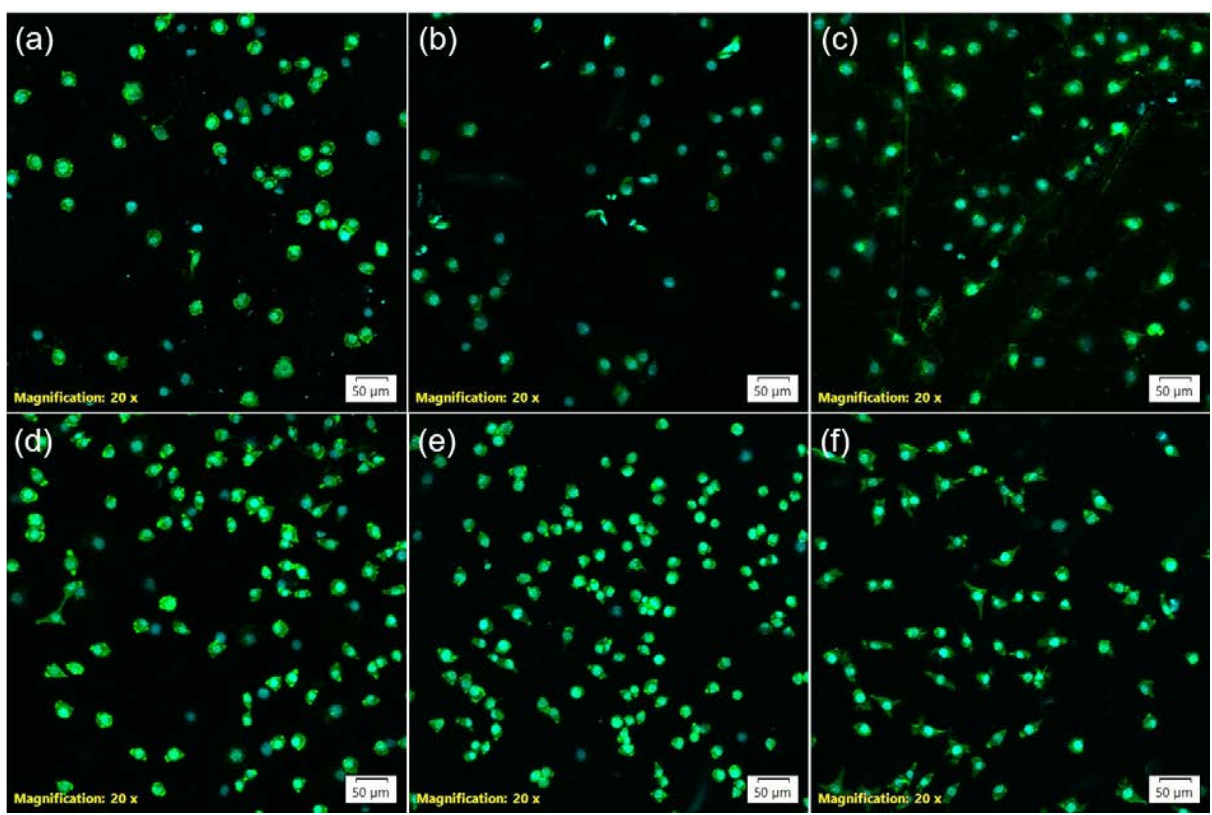

**Figure 7.** Confocal microscopy images of the interaction between L929 cells and the surface of polymer films for 24 h. (a) PCL, (b)PCE$_{0.6k}$C, (c)PCE$_{2k}$C, (d)PCE$_{6k}$C, (e)PCE$_{20k}$C and (f)PCE$_{35k}$C.

### 3.6 Application in the 3D-printed scaffold of tissue engineering

The 3D printed PCEC copolymer scaffolds were fabricated by pellet-based screw-assisted 3D printers. It is highly tolerant to the shape of the material, such as powders, pellets, or gels. Compared with the commonly used fused deposition modeling (FDM) system, it reduces the cumbersome process of making uniform filaments, which is simpler and more efficient for manufacturing. The printing parameters of 3D printing were greatly influenced by the rheological behavior, especially the extrusion flow rate, which is directly related to the viscosity. As shown in **Figure 3d**, with the increase of PEG content, the PCEC copolymers show increased complex viscosities, which indicates higher extrusion rates are required to obtain a smooth printing process.

However, due to the shear thinning behavior, even though PCE$_{20k}$C and PCE$_{35k}$C exhibit high viscosities, they can be extruded at low extrusion flow rates, improving printing efficiency and reducing nozzle clogging issues.

PCL and PCEC copolymers can be all printed into scaffolds with precise geometry and high pore connectivity. Taking PCE$_{20k}$C as an example, 3D printed scaffolds with different sizes, number of layers, and porosity were displayed in **Figure 8**. Because of the flexibility of PCE$_{20k}$C, the 2-layer sheet shows excellent bending performance, which can be curled or bent multiple times without damage after recovery, showing potential for applications in joint tissue engineering. In addition, scaffolds with smaller dimensions and more complex pore structures can also be obtained. It can be seen from the optical microscopy images that, the printed filaments of the scaffolds are smooth and uniform, cylindrical in shape, with a diameter of ~ 500 μm close to the designed size. The printed scaffolds also show a well-defined and interconnected porous network structure, and their sizes were also close to the designed 1000×1000 μm and 500×500 μm, respectively. Besides, no collapse of filaments and pore clogging are observed on the top surface of the scaffold, indicating its good printability. These results confirm the usability of PCEC copolymers in 3D printed scaffolds, and the properties of printed scaffolds will be explored in further studies.

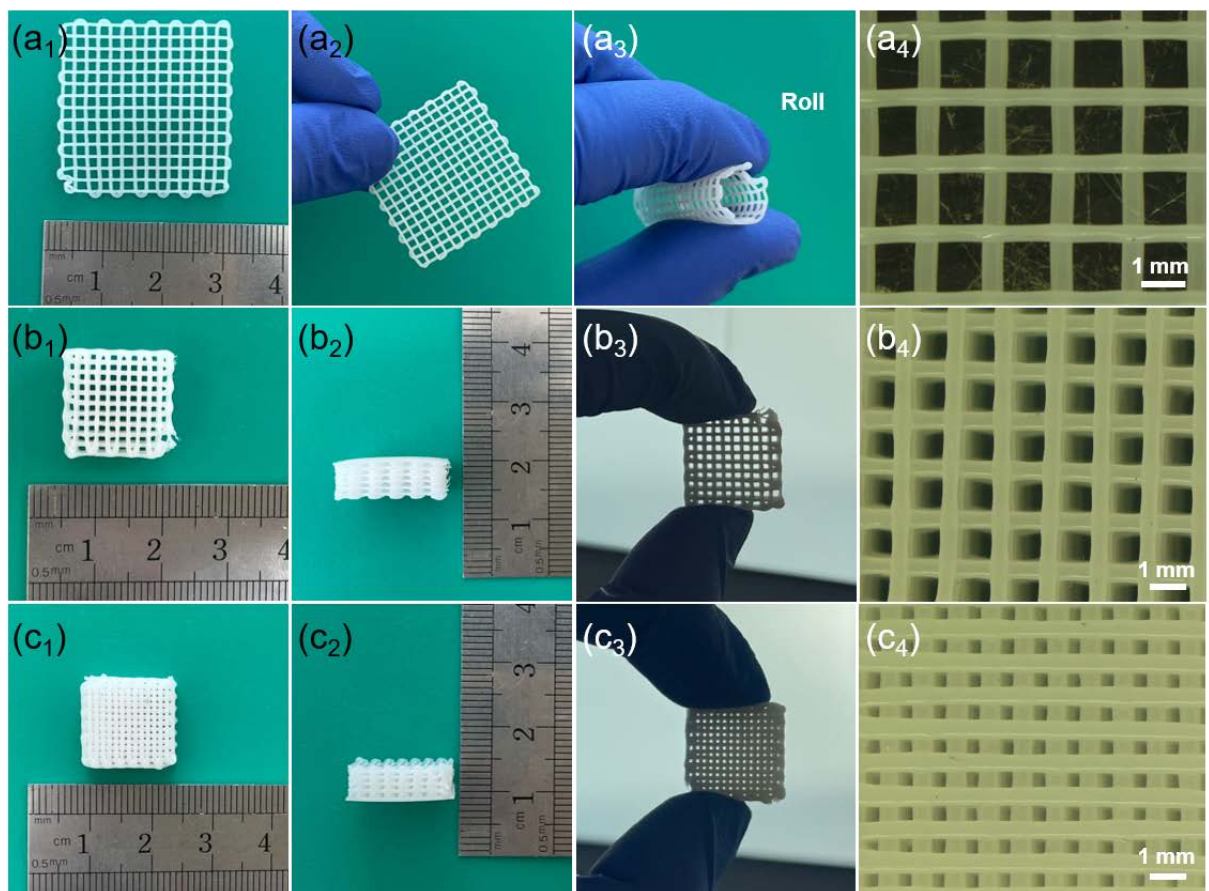

**Figure 8.** Digital photos and optical microscopy images of the 3D printed PCE$_{20k}$C scaffolds. (a) a 2-layered sheet, (b) a 10-layered scaffold with 1000 μm porosity and (c) a 10-layered scaffold with 500 μm porosity

## 4. Conclusions

Biodegradable and biocompatible PCEC copolymers with tunable mechanical properties and degradation rate were synthesized by using macroinitiators ranging from low molecular weight PEG (0.6k g/mol) to high molecular weight PEG (35k g/mol). Three different crystallization regimes were found depending on PCL/PEG block ratio which significantly affected the mechanical properties. The elastic modulus of the copolymers could be tuned between 338 - 705 MPa, close to the value of cancellous bone, tendon, and collateral ligament, while the elongation to fracture was always > 500%. The surface hydrophilicity and degradation rate were directly related to the PEG content. PCEC copolymers with higher PEG content showed higher hydrophilicity, faster degradation rate, and improved cell attachment. In fact, PCE$_{20k}$C and PCE$_{35k}$C copolymers exhibited extremely fast apparent degradation rates in 5M NaOH solution. Moreover, all PCEC copolymers could be manufactured into 3D scaffolds with controlled microstructure, particularly PCE$_{20k}$C and PCE$_{35k}$C which exhibit excellent shear-thinning rheological properties. These results show that is possible to tailor the mechanical properties, degradation rate and hydrophilicity of PCEC copolymers to adapt the properties of the copolymers to different native tissues, and they also provide a strategy for the design and selection of other copolymers for biomedical applications.

## Author contributions

Yu-Yao Liu: conceptualization, methodology, validation, formal analysis, investigation and writing - original draft. Juan Pedro Fernandez Blazquez: investigation, validation and formal analysis. Guang-Zhong Yin: investigation and validation. De-Yi Wang: investigation and validation. Javier Llorca: conceptualization, supervision, writing – review and editing and funding acquisition. Monica Echeverry-Rendón: conceptualization, supervision, and writing – review and editing.

## Conflicts of interest

There are no conflicts to declare.

# Acknowledgements

This investigation was supported by the from the Spanish Research Agency through the grant PID2021-124389OB-C21. Ms. Y. Liu acknowledges the support from the China Scholarship Council (CSC No. 202106990026). In addition, the technical assistance of Ms. J de la Vega and Wen-Qiang Yuan are gratefully acknowledged.

# References

[1] X. Zhang, Z. Li, P. Yang, G. Duan, X. Liu, Z. Gu, Y. Li, Polyphenol scaffolds in tissue engineering, Mater. Horiz. 8 (2021), 145–167. https://doi.org/10.1039/d0mh01317j.

[2] Tissue engineering, Nat. Biotechnol. 18 (2000), IT56–IT58. https://doi.org/10.1038/80103.

[3] J. Malda, J. Visser, F.P. Melchels, T. Jüngst, W.E. Hennink, W.J.A. Dhert, J. Groll, D.W. Hutmacher, 25th anniversary article: Engineering hydrogels for biofabrication, Adv. Mater. 25 (2013), 5011–5028. https://doi.org/10.1002/adma.201302042.

[4] B.D. Ulery, L.S. Nair, C.T. Laurencin, Biomedical applications of biodegradable polymers, J. Polym. Sci. B: Polym. Phys. 49 (2011), 832–864. https://doi.org/10.1002/polb.22259.

[5] S.H. Hsu, K.C. Hung, C.W. Chen, Biodegradable polymer scaffolds, J. Mater. Chem. B., 4 (2016), 7493–7505. https://doi.org/10.1039/c6tb02176j.

[6] C. Ning, P. Li, C. Gao, L. Fu, Z. Liao, G. Tian, H. Yin, M. Li, X. Sui, Z. Yuan, S. Liu, Q. Guo, Recent advances in tendon tissue engineering strategy, Front. Bioeng. Biotechnol. 11 (2023),. https://doi.org/10.3389/fbioe.2023.1115312.

[7] M. Gasik, A. Zühlke, A.M. Haaparanta, V. Muhonen, K. Laine, Y. Bilotsky, M. Kellomäki, I. Kiviranta, The importance of controlled mismatch of biomechanical compliances of implantable scaffolds and native tissue for articular cartilage regeneration, Front Bioeng Biotechnol. 6 (2018),187. https://doi.org/10.3389/fbioe.2018.00187.

[8] B.S. Heidari, P. Chen, R. Ruan, S.M. Davachi, H. Al-Salami, E. De Juan Pardo, M. Zheng, B. Doyle, B.S. Heidari, A novel biocompatible polymeric blend for applications requiring high toughness and tailored degradation rate, J. Mater. Chem. B. 9 (2021), 2532–2546. https://doi.org/10.1039/d0tb02971h.

[9] E.J. Go, E.Y. Kang, S.K. Lee, S. Park, J.H. Kim, W. Park, I.H. Kim, B. Choi, D.K. Han, An osteoconductive PLGA scaffold with bioactive β-TCP and anti-inflammatory Mg(OH)2 to improve: In vivo bone regeneration, Biomater. Sci. 8 (2020), 937–948. https://doi.org/10.1039/c9bm01864f.

[10] H. Ben Amara, D.C. Martinez, F.A. Shah, A.J. Loo, L. Emanuelsson, B. Norlindh, R. Willumeit-Römer, T. Plocinski, W. Swieszkowski, A. Palmquist, O. Omar, P. Thomsen, Magnesium implant degradation provides immunomodulatory and proangiogenic effects and attenuates peri-implant fibrosis in soft tissues, Bioact. Mater. 26 (2023), 353–369. https://doi.org/10.1016/j.bioactmat.2023.02.014.

[11] L. Zhao, X. Pei, L. Jiang, C. Hu, J. Sun, F. Xing, C. Zhou, Y. Fan, X. Zhang, Bionic design and 3D printing of porous titanium alloy scaffolds for bone tissue repair, Compos. B Eng. 162 (2019), 154–161. https://doi.org/10.1016/j.compositesb.2018.10.094.

[12] Z.U. Arif, M.Y. Khalid, R. Noroozi, A. Sadeghianmaryan, M. Jalalvand, M. Hossain, Recent advances in 3D-printed polylactide and polycaprolactone-based biomaterials for

tissue engineering applications, Int. J. Biol. Macromol. 218 (2022), 930–968. https://doi.org/https://doi.org/10.1016/j.ijbiomac.2022.07.140.

[13]  H. Ma, C. Feng, J. Chang, C. Wu, 3D-printed bioceramic scaffolds: From bone tissue engineering to tumor therapy, Acta Biomater. 79 (2018), 37–59. https://doi.org/10.1016/j.actbio.2018.08.026.

[14]  A.V. Do, B. Khorsand, S.M. Geary, A.K. Salem, 3D Printing of Scaffolds for Tissue Regeneration Applications, Adv. Healthc. Mater. 4 (2015), 1742–1762. https://doi.org/10.1002/adhm.201500168.

[15]  A. Kirillova, T.R. Yeazel, D. Asheghali, S.R. Petersen, S. Dort, K. Gall, M.L. Becker, Fabrication of Biomedical Scaffolds Using Biodegradable Polymers, Chem. Rev. 121 (2021), 11238–11304. https://doi.org/10.1021/acs.chemrev.0c01200.

[16]  A.V. Do, B. Khorsand, S.M. Geary, A.K. Salem, 3D Printing of Scaffolds for Tissue Regeneration Applications, Adv. Healthc. Mater. 4 (2015), 1742–1762. https://doi.org/10.1002/adhm.201500168.

[17]  J. Zhang, D. Tong, H. Song, R. Ruan, Y. Sun, Y. Lin, J. Wang, L. Hou, J. Dai, J. Ding, H. Yang, Osteoimmunity-Regulating Biomimetically Hierarchical Scaffold for Augmented Bone Regeneration, Adv. Mater. 34 (2022), 2202044. https://doi.org/10.1002/adma.202202044.

[18]  C. Wang, W. Huang, Y. Zhou, L. He, Z. He, Z. Chen, X. He, S. Tian, J. Liao, B. Lu, Y. Wei, M. Wang, 3D printing of bone tissue engineering scaffolds, Bioact. Mater. 5 (2020), 82–91. https://doi.org/10.1016/j.bioactmat.2020.01.004.

[19]  S.C. Daminabo, S. Goel, S.A. Grammatikos, H.Y. Nezhad, V.K. Thakur, Fused deposition modeling-based additive manufacturing (3D printing): techniques for polymer material systems, Mater. Today Chem. 16 (2020), 100248. https://doi.org/10.1016/j.mtchem.2020.100248.

[20]  S.C. Altıparmak, V.A. Yardley, Z. Shi, J. Lin, Extrusion-based additive manufacturing technologies: State of the art and future perspectives, J. Manuf. Process. 83 (2022), 607–636. https://doi.org/10.1016/j.jmapro.2022.09.032.

[21]  B. Guo, P.X. Ma, Synthetic biodegradable functional polymers for tissue engineering: A brief review, Sci. China Chem. 57 (2014), 490–500. https://doi.org/10.1007/s11426-014-5086-y.

[22]  C. Xu, Y. Hong, Rational design of biodegradable thermoplastic polyurethanes for tissue repair, Bioact. Mater. 15 (2022), 250–271. https://doi.org/10.1016/j.bioactmat.2021.11.029.

[23]  X. Liu, J.M. Holzwarth, P.X. Ma, Functionalized Synthetic Biodegradable Polymer Scaffolds for Tissue Engineering, Macromol. Biosci. 12 (2012), 911–919. https://doi.org/10.1002/mabi.201100466.

[24] J. Kundu, F. Pati, Y. Hun Jeong, D.W. Cho, Biomaterials for Biofabrication of 3D Tissue Scaffolds, in: Biofabrication: Micro- and Nano-Fabrication, Printing, Patterning and Assemblies, Elsevier Inc., 2(2013), 23–46. https://doi.org/10.1016/B978-1-4557-2852-7.00002-0.

[25] S. Stratton, N.B. Shelke, K. Hoshino, S. Rudraiah, S.G. Kumbar, Bioactive polymeric scaffolds for tissue engineering, Bioact. Mater. 1 (2016), 93–108. https://doi.org/10.1016/j.bioactmat.2016.11.001.

[26] D. Zhao, T. Zhu, J. Li, L. Cui, Z. Zhang, X. Zhuang, J. Ding, Poly(lactic-co-glycolic acid)-based composite bone-substitute materials, Bioact. Mater. 6 (2021), 346–360. https://doi.org/10.1016/j.bioactmat.2020.08.016.

[27] Y. Hou, W. Wang, P. Bartolo, Investigation of polycaprolactone for bone tissue engineering scaffolds: In vitro degradation and biological studies, Mater. Des. 216 (2022), 110582. https://doi.org/10.1016/j.matdes.2022.110582.

[28] S. Vijayavenkataraman, S. Thaharah, S. Zhang, W.F. Lu, J.Y.H. Fuh, Electrohydrodynamic jet 3D-printed PCL/PAA conductive scaffolds with tunable biodegradability as nerve guide conduits (NGCs) for peripheral nerve injury repair, Mater. Des. 162 (2019), 171–184. https://doi.org/10.1016/j.matdes.2018.11.044.

[29] S. Wang, R. Gu, F. Wang, X. Zhao, F. Yang, Y. Xu, F. Yan, Y. Zhu, D. Xia, Y. Liu, 3D-Printed PCL/Zn scaffolds for bone regeneration with a dose-dependent effect on osteogenesis and osteoclastogenesis, Mater. Today Bio. 13 (2022), 100202. https://doi.org/10.1016/j.mtbio.2021.100202.

[30] S. Murab, S.M.S. Gruber, C.Y.J. Lin, P. Whitlock, Elucidation of bio-inspired hydroxyapatie crystallization on oxygen-plasma modified 3D printed poly-caprolactone scaffolds, Mater. Sci. Eng. C 109 (2020), 110529. https://doi.org/10.1016/j.msec.2019.110529.

[31] Q. Dong, M. Zhang, X. Zhou, Y. Shao, J. Li, L. Wang, C. Chu, F. Xue, Q. Yao, J. Bai, 3D-printed Mg-incorporated PCL-based scaffolds: A promising approach for bone healing, Mater. Sci. Eng. C 129 (2021), 112372. https://doi.org/10.1016/j.msec.2021.112372.

[32] C. Vyas, J. Zhang, Ø. Øvrebø, B. Huang, I. Roberts, M. Setty, B. Allardyce, H. Haugen, R. Rajkhowa, P. Bartolo, 3D printing of silk microparticle reinforced polycaprolactone scaffolds for tissue engineering applications, Mater. Sci. Eng. C 118 (2021), 111433. https://doi.org/10.1016/j.msec.2020.111433.

[33] H. Ehtesabi, F. Massah, Improvement of hydrophilicity and cell attachment of polycaprolactone scaffolds using green synthesized carbon dots, Mater. Today Sus. 13 (2021), 100075. https://doi.org/10.1016/j.mtsust.2021.100075.

[34] H. Maleki-Ghaleh, M. Hossein Siadati, A. Fallah, A. Zarrabi, F. Afghah, B. Koc, E. Dalir Abdolahinia, Y. Omidi, J. Barar, A. Akbari-Fakhrabadi, Y. Beygi-Khosrowshahi, K. Adibkia, Effect of zinc-doped hydroxyapatite/graphene nanocomposite on the

physicochemical properties and osteogenesis differentiation of 3D-printed polycaprolactone scaffolds for bone tissue engineering, Chem. Eng. J. 426 (2021), 131321. https://doi.org/10.1016/j.cej.2021.131321.

[35] C. Cao, P. Huang, A. Prasopthum, A.J. Parsons, F. Ai, J. Yang, Characterisation of bone regeneration in 3D printed ductile PCL/PEG/hydroxyapatite scaffolds with high ceramic microparticle concentrations, Biomater. Sci. 10 (2022), 138–152. https://doi.org/10.1039/d1bm01645h.

[36] P.Y. Ni, Q.X. Ding, M. Fan, J.F. Liao, Z.Y. Qian, J.C. Luo, X.Q. Li, F. Luo, Z.M. Yang, Y.Q. Wei, Injectable thermosensitive PEG-PCL-PEG hydrogel/acellular bone matrix composite for bone regeneration in cranial defects, Biomaterials 35 (2014), 236–248. https://doi.org/10.1016/j.biomaterials.2013.10.016.

[37] C.P. Jiang, J.R. Huang, M.F. Hsieh, Fabrication of synthesized PCL-PEG-PCL tissue engineering scaffolds using an air pressure-aided deposition system, Rapid Prototyp. J. 17 (2011), 288–297. https://doi.org/10.1108/13552541111138414.

[38] N. Fu, J. Liao, S. Lin, K. Sun, T. Tian, B. Zhu, Y. Lin, PCL-PEG-PCL film promotes cartilage regeneration in vivo, Cell Prolif. 49 (2016), 729–739. https://doi.org/10.1111/cpr.12295.

[39] M. Nagata, I. Kitazima, Photocurable biodegradable poly(ε-caprolactone)/poly(ethylene glycol) multiblock copolymers showing shape-memory properties, Colloid Polym. Sci. 284 (2006), 380–386. https://doi.org/10.1007/s00396-005-1393-3.

[40] J.Z. Bei, J.M. Li, Z.F. Wang, J.C. Le, S.G. Wang, Polycaprolactone-poly(ethylene-glycol) block copolymer. IV: Biodegradation behavior in vitro and in vivo, Polym. Adv. Technol. 8 (1997), 693–696. https://doi.org/10.1002/(SICI)1099-1581(199711)8:11<693::AID-PAT702>3.0.CO;2-B.

[41] J. Sun, J. Yang, J. Ding, Controlled Synthesis of Polymers†, Chin. J. Chem. 41 (2023), 1235–1248. https://doi.org/10.1002/cjoc.202200850.

[42] G. Yin, D. Zhao, X. Wang, Y. Ren, L. Zhang, X. Wu, S. Nie, Q. Li, Bio-compatible poly(ester-urethane)s based on PEG-PCL-PLLA copolymer with tunable crystallization and bio-degradation properties, RSC Adv. 5 (2015), 79070–79080. https://doi.org/10.1039/c5ra15531b.

[43] R. Il Kim, G. Lee, J.H. Lee, J.J. Park, A.S. Lee, S.S. Hwang, Structure-Property Relationships of 3D-Printable Chain-Extended Block Copolymers with Tunable Elasticity and Biodegradability, ACS Appl. Polym. Mater. 3 (2021), 4708–4716. https://doi.org/10.1021/acsapm.1c00860.

[44] Y. Huang, L. Li, G. Li, An enzyme-catalysed access to amphiphilic triblock copolymer of PCL-b-PEG-b-PCL: synthesis, characterization and self-assembly properties, Des. Monomers Polym. 18 (2015), 799–806. https://doi.org/10.1080/15685551.2015.1078113.

[45] K.C. Ang, K.F. Leong, C.K. Chua, M. Chandrasekaran, Compressive properties and degradability of poly(ε-caprolatone)/ hydroxyapatite composites under accelerated hydrolytic degradation, J. Biomed. Mater. Res. A 80 (2007), 655–660. https://doi.org/10.1002/jbm.a.30996.

[46] M. Bartnikowski, T.R. Dargaville, S. Ivanovski, D.W. Hutmacher, Degradation mechanisms of polycaprolactone in the context of chemistry, geometry and environment, Prog. Polym. Sci. 96 (2019), 1–20. https://doi.org/10.1016/j.progpolymsci.2019.05.004.

[47] F. Von Burkersroda, L. Schedl, A.G. Opferich, Why degradable polymers undergo surface erosion or bulk erosion, Biomaterials 23 (2002), 4221–4231. https://doi.org/10.1016/S0142-9612(02)00170-9.